\documentclass[aps,showpacs]{revtex4}
\usepackage{graphicx}
\usepackage{color}
\usepackage[T1]{fontenc}

\RequirePackage{amsfonts,amssymb,amsmath}

\begin{document}
\title{Quantum singularities in FRW universe revisited}
\author{Patricio S. Letelier} 
 \email{e-mail: letelier@ime.unicamp.br} 
\author{Jo\~ao Paulo M. Pitelli} 
\email{e-mail:pitelli@ime.unicamp.br}
\affiliation{
Departamento de Matem\'atica Aplicada-IMECC,
Universidade Estadual de Campinas,
13081-970 Campinas,  Sao Paulo, Brazil}

\begin{abstract}
The components of the Riemann tensor in the tetrad basis are quantized and, through the Einstein equation, we find the local expectation value in the ontological interpretation of quantum mechanics of the energy density and pressure of a perfect fluid with equation of state $p=\frac{1}{3}\rho$ in the flat Friedmann-Robertson-Walker quantum cosmological model. The quantum behavior of the equation of state and energy conditions are then studied and it is shown that the later is violated since the singularity is removed with the introduction of quantum cosmology, but in the classical limit both the equation of state and the energy conditions behave as in the classical model. We also calculate the expectation value of the scale factor  for several wave packets in the many-worlds interpretation in order to show the independence of the non singular character of the quantum cosmological model with respect to the wave packet representing the wave function of the Universe. It is also shown that, with the introduction of non-normalizable wave packets, solutions of the Wheeler-DeWitt equation, the singular character of the scale factor, can be recovered in the ontological interpretation.
\end{abstract}

\pacs{ 98.80.Qc,   04.60.Ds, 04.20.Dw}
\maketitle

\section{Introduction}
One of the main problems of modern cosmology is the presence of singularities in cosmological models. Classical singularities in general relativity are indicated by incomplete geodesics or incomplete paths of bounded acceleration \cite{hawking}. There are three types of singularities \cite{helliwell2,konkowski}: the quasi regular singularity, where no observer sees any physical quantities diverging even if its world line reaches the singularity (for example the singularity in the spacetime of a cosmic string); the scalar curvature singularity, where every observer near the singularity sees physical quantities diverging [for example the singularity in the Schwarchild spacetime or the big-bang singularity in Friedmann-Robertson-Walker (FRW) cosmology]; the non scalar curvature singularity, where there are some curves in which the observers experience unbounded tidal forces (whimper cosmologies are a good example). It was shown that under very reasonable conditions (the energy conditions) singularities are always present in cosmological models \cite{hawking}. Since general relativity cannot escape this burden, we hope that a complete theory of quantum gravity will overcome this situation, teaching us how to deal with such spacetime near the singularities or excluding the singularities at all. While such a theory does not exist, there are many attempts to incorporate quantum mechanics into general relativity. One of the first attempts to do this was quantum cosmology \cite{dewitt,haliwell}. One of the major problem of quantum cosmology is the absence of a natural time variable, since the Wheeler-DeWitt equation is a second order functional differential equation in the superspace, and there is no first order functional differential to play the role of time. The introduction of matter fields may overcome this situation, since the evolution of a dynamical parameter of the matter field can recover the notion of time. With the introduction of a perfect fluid in Schutz's formalism \cite{schutz1,schutz2}, we can recover a Schr\"odinger-like equation in the minisuperspace in which the momentum associated with the dynamical degree of freedom of the fluid appears linearly in the equation. 

Another problem of quantum cosmology is the interpretation of the wave function of the Universe. Since the Universe includes everything, the probabilistic interpretation becomes impossible, since it assumes there is a fundamental process of measure outside the quantum world, in a classical domain \cite{pinto-neto}. Two of the most common alternative interpretations to the wave function in quantum cosmology are the {\it many-worlds} \cite{everett} and the {\it de Broglie-Bohm} \cite{bohm} interpretation of quantum mechanics. In the former, all the possibilities in the splitting are actually realized, and there is one observer in each branch with the knowledge of the correspondent eigenvalue, and these branches do not interfere. Every time an experiment is performed, every one of the outcomes are obtained, each in a different world.  In the later, a trajectory in the phase space for each dynamical variable is supposed to exist independently of any measure. In a measurement, this trajectory enters in one of the branches, depending on the initial condition, which is unknown.

In this paper we use the machinery of quantum cosmology and the de Broglie-Bohm interpretation of quantum mechanics in order to study the energy conditions in the quantum FRW universe with a perfect fluid in the radiation dominated era. We hope that the energy conditions will be violated at the quantum domain since it was shown that the big-bang singularity in the FRW universe is removed with the introduction of quantum mechanics \cite{monerat}. We also study the behavior of the scale factor in the many-worlds interpretation for several wave packets to see if the exclusion of the big-bang singularity is not particular to the simple wave packets found in the literature so far.

In what follows we proceed in the following manner: In Sec. \ref{section 2}, we present, for easy reference,  a brief summary of quantum cosmology in the flat FRW universe in the radiation dominated era. In Sec. \ref{section 3} we quantize the components of the Riemann tensor in the tetrad basis and study the energy conditions and the equation of state for the quantum cosmological model of the Universe in order to show the consistency of the model. In Sec. \ref{section 4}, we find the expectation value  of the scale factor for several wave packets in  the many-worlds interpretation  to show the independence of the results found previously in the literature with respect to the particular wave packet. In Sec. \ref{section 5}, we show that with the introduction of  nonnormalizable wave packets, the classical singular behavior can be recovered. Finally, in Sec. \ref{section 6}, we discuss the main results presented in this work.

\section{Wheeler-DeWitt equation for the flat FRW universe with a perfect fluid}
\label{section 2}
The action for general relativity is given by
\begin{equation}
S_{G}=\int_{\mathcal{M}}{d^{4}x\sqrt{-g}R}+2\int_{\partial \mathcal{M}}{d^3x\sqrt{h}h_{ab}K^{ab}},
\label{action for general relativity}
\end{equation}
where $h_{ab}$ is the induced metric over the boundary $\partial \mathcal{M}$ of the four-dimensional manifold $\mathcal{M}$ and $K^{ab}$ is the extrinsic curvature of the hypersurface $\partial \mathcal{M}$, that is, the curvature of $\partial \mathcal{M}$ with respect to $\mathcal{M}$.

 Schutz \cite{schutz2} showed that for a perfect fluid with four velocity expressed in terms of five potentials
\begin{equation}
U_{\nu}=\mu^{-1}(\phi_{,\nu}+\alpha\beta_{,\nu}+\theta S_{,\nu}),
\label{representation of velocity}
\end{equation}
where $\mu$ is the specific enthalpy, and respecting the normalization condition
\begin{equation}
U_\nu U^\nu=-1, 
\label{normalization of velocity}
\end{equation}
the action is given by
\begin{equation}
S_{f}=\int_{\mathcal{M}}{d^4x\sqrt{-g}p},
\label{action general relativity}
\end{equation}
where $p$ is the  fluid pressure, which is linked to the energy density by the state  equation,  $p=\alpha \rho$.

The super-Hamiltonian for the total action,
\begin{equation}
S=S_{G}+S_{f},
\label{total action}
\end{equation}
for the flat FRW universe,
\begin{equation}
ds^2=-N(t)^2dt^2+a^2(t)(dr^2+r^2d\Omega^2),
\label{FRW}
\end{equation}
with matter represented by perfect fluid with equation of state $p=\frac{1}{3}\rho$ is given by \cite{lapichinskii,alvarenga}
\begin{equation}
\mathcal{H}=\frac{p_a^2}{24}-p_T,
\label{super-Hamiltonian}
\end{equation}
where $p_a=-12\frac{da}{dT}a/N$ is the momentum conjugated to the scale factor $a(t)$ and $p_T$ is the momentum conjugated to the dynamical degree of freedom of the fluid. In fact, the super-Hamiltonian (\ref{super-Hamiltonian}) is a constraint of  the theory. By following the Dirac approach \cite{dirac} for quantization of Hamiltonian systems with constraints, imposing
\begin{equation}
p_a=-i\frac{\partial}{\partial a};\;\;\;\;\;p_T=-i\frac{\partial}{\partial T}
\label{canonical quantization}
\end{equation}
and demanding that the super-Hamiltonian constraint annihilate the wave function we find \cite{alvarenga}
\begin{equation}
\frac{\partial^2\Psi}{\partial a^2}+24i\frac{\partial \Psi}{\partial t}=0,
\label{wheeler-dewitt equation}
\end{equation}
where $t=-T$ is the time coordinate in the conformal-time gauge $N=a$.

The internal product between two wave functions is defined by
\begin{equation}
\left<\Psi|\Phi\right>=\int_{0}^{\infty}{\Psi(a,t)^{\ast}\Phi(a,t)da},
\end{equation}
so that the condition for self-adjointness of the operator $\hat{H}=-\frac{\partial^2}{\partial a^2}$ is given by \cite{lemos}
\begin{equation}
\Psi'(0,t)=\beta \Psi(0,t),\;\;\;\beta\in\mathbb{R}.
\label{boundary conditions}
\end{equation}

\section{Energy conditions}
\label{section 3}
Equation (\ref{wheeler-dewitt equation}) is analogous to the Schr\"odinger equation for a free-particle. Its Green function in the usual $L^2(-\infty,\infty)$ space is given by \cite{lemos}
\begin{equation}
G(a,a',t)=\left(\frac{6}{\pi i t}\right)^{1/2}\exp{\left[\frac{6i}{t}(a-a')^2\right]}.
\label{green function}
\end{equation}
With the   propagator above  we can evolve any given initial wave packet for the universe. We choose a normalized wave packet satisfying the boundary condition $\Psi'(0,t)=0$ ($\beta=0$),
\begin{equation}
\Psi(a,0)=\left(\frac{8\sigma}{\pi}\right)^{1/4}e^{-\sigma a^2}.
\label{initial wave packet}
\end{equation}
The Green function for the chosen boundary condition is given by
\begin{equation}
G^{\beta=0}(a,a',t)=G(a,a',t)+G(a,-a',t),
\label{green function beta=0}
\end{equation}
so that
\begin{subequations}\begin{align}
\Psi(a,t)&=\int_{0}^{\infty}{G^{\beta=0}(a,a',t)\Psi(a',0)da'}=\int_{-\infty}^{\infty}{G(a,a',t)\Psi(a',0)da'} \label{eqa}\\&=\left(\frac{8\sigma}{\pi}\right)^{1/4}\left(\frac{6}{\sigma t-6i}\right)^{1/2}\exp{\left(\frac{6i\sigma a^2}{\sigma t -6 i}\right)}.
\label{wave packet with green function}\end{align}
\end{subequations}
 In the first line of Eq. (\ref{eqa}) we used the fact that $\Psi(a',0)$ is an even function of $a'$ so that the first integral above is mathematically equivalent to the second one.

The analysis of this section will be made using the above wave packet, which satisfies the Neumann boundary condition $\Psi'(0,t)=0$. If we chose a wave packet satisfying the Dirichlet boundary condition $\Psi(0,t)=0$, the result is the same, since in both cases the quantum Universe is nonsingular (the wave packet satisfying the Dirichlet boundary condition is more complicated \cite{lemos}). In fact, the expectation value of the scale factor in the many-worlds interpretation for the wave packet satisfying the Dirichlet boundary condition is twice the value of that satisfying the Neumann boundary condition \cite{lemos}. The Bohmian trajectory of the scale factor is the same on both cases. No matter what choice we make, the Hamiltonian operator is self-adjoint so that the evolution of the wave packet is unitary.

In order to use the de Broglie-Bohm interpretation, we need to write the wave function of the universe in its polar form
\begin{equation}
\Psi=\Theta e^{iS},
\label{general polar form}
\end{equation}
where $\Theta$ and $S$ are real functions. In our case,
\begin{equation}\begin{aligned}
&S=\frac{6\sigma^2 t a^2}{\sigma^2t^2+36}+f_{0}(t),\\
&\Theta=g_{0}(t)\exp\left(-\frac{36\sigma a^2}{\sigma^2t^2+36}\right),
\label{theta and S}\end{aligned}
\end{equation}
where $f_{0}(t)$ and $g_0(t)$ will not matter in what comes.

Now we can calculate the Bohmian trajectories for the scale factor $a(t)$ by the use of the equation \cite{bohm,holland}
\begin{equation}
p_a=\frac{\partial S}{\partial a}.
\label{momentum bohmian}
\end{equation}
So we have
\begin{equation}
12\dot{a}=\frac{12\sigma^2 t a}{\sigma^2t^2+36}\Rightarrow a(t)=a_0\sqrt{36+\sigma^2t^2},
\label{bohmian trajectory}
\end{equation}
where $a_0$ is an integration constant.

For the metric (\ref{FRW}) we have as a basis for the cotangent space the following 1-forms
\begin{equation}
\omega^{\hat{t}}=N(t)dt;\;\;\;\;\;\omega^{\hat{r}}=a(t)dr;\;\;\;\;\;\omega^{\hat{\theta}}=a(t)rd\theta;\;\;\;\;\;\omega^{\hat{\phi}}=a(t)r\sin{\theta}d\phi.
\label{1-forms}
\end{equation}
In this basis the metric can be written as 
\begin{equation}
ds^2=-(\omega^{\hat{t}})^{2}+(\omega^{\hat{r}})^{2}+(\omega^{\hat{\theta}})^{2}+(\omega^{\hat{\phi}})^{2}=\eta_{\hat{a}\hat{b}}\omega^{\hat{a}}\omega^{\hat{b}}.
\label{metric in the tetrad basis}
\end{equation}

There are only two independent components of the curvature tensor in the tetrad basis. They are
\begin{equation}\begin{aligned}
&R_{\hat{t}\hat{r}\hat{t}\hat{r}}=R_{\hat{t}\hat{\theta}\hat{t}\hat{\theta}}=R_{\hat{t}\hat{\phi}\hat{t}\hat{\phi}}=-\frac{\ddot{a}}{a^3}+\frac{\dot{a}^2}{a^4},\\
&R_{\hat{r}\hat{\theta}\hat{r}\hat{\theta}}=R_{\hat{r}\hat{\phi}\hat{r}\hat{\phi}}=R_{\hat{\theta}\hat{\phi}\hat{\theta}\hat{\phi}}=\frac{\dot{a}^2}{a^4}.
\label{curvature components}\end{aligned}
\end{equation}
By the relation $p_a=12\dot{a}$, we have
\begin{equation}\begin{aligned}
&R_{\hat{t}\hat{r}\hat{t}\hat{r}}=-\frac{\dot{p}_a}{12a^3}+\frac{p_a^2}{144a^4},\\
&R_{\hat{r}\hat{\theta}\hat{r}\hat{\theta}}=\frac{p_a^2}{144a^4}.
\label{components momentum}\end{aligned}
\end{equation}

Here we promote the components of the curvature tensor to the condition of quantum operators. Then we can take the local expectation value of each component through the relation \cite{holland}
\begin{equation}
\left<R_{\hat{a}\hat{b}\hat{c}\hat{d}}\right>_L=\text{Re}\left(\frac{\Psi^{\ast}\hat{R}_{\hat{a}\hat{b}\hat{c}\hat{d}}\Psi}{\Psi^{\ast}\Psi}\right).
\label{expectation value}
\end{equation}

Equation (\ref{components momentum}) mixes the coordinate $a$ with its respective momentum $p_a$. So we face a factor ordering problem since $a$ and $p_a$ do not commute. Here, following the choice of Ref. \cite{monerat}, we chose the Weyl ordering \cite{lee}. Weyl ordering is a kind of symmetrization procedure that take all possible orders of the $a$'s and the $p_a$'s and then divides the result by the number of terms in the final expression.
In the case of an operator of the form $f(a)p_a^2$, Weyl ordering corresponds to the following expression
\begin{equation}
\left(f(a)\hat{p}_a^2\right)_W=\frac{1}{4}\left[f(a)\hat{p}_a^2+2\hat{p}_af(a)\hat{p}_a+\hat{p}_a^2f(a)\right].
\label{weyl ordering}
\end{equation}
Using the relations

\begin{equation}\begin{aligned}
&\left[a^{-4},\hat{p}_a\right]=-4ia^{-5},\\
&\left[a^{-4},\hat{p}_a^2\right]=-8ia^{-5}p_a+20a^{-6},
\label{comutation relations}\end{aligned}
\end{equation}
we have
\begin{equation}\begin{aligned}
&\hat{p}_aa^{-4}=a^{-4}\hat{p}_a+4ia^{-5},\\
&\hat{p}_a^2a^{-4}=a^{-4}\hat{p}_a^2+8ia^{-5}\hat{p}_a-20a^{-6}.
\label{momentum in the front}\end{aligned}
\end{equation}

Then, after Weyl ordering operation, the operator $a^{-4}\hat{p}_a$ becomes
\begin{equation}
\left(\frac{\hat{p}_a^2}{a^4}\right)_W=a^{-4}\hat{p}_a^2+4ia^{-5}\hat{p}_a-5a^{-6}.
\label{weyl operator}
\end{equation}

The components of the curvature tensor can be written as
\begin{equation}\begin{aligned}
&\left(\hat{R}_{\hat{t}\hat{r}\hat{t}\hat{r}}\right)_W=-\frac{\dot{p}_a}{12a^3}+\frac{1}{144}\left(a^{-4}\hat{p}_a^2+4ia^{-5}\hat{p}_a-5a^{-6}\right),\\
&\left(\hat{R}_{\hat{r}\hat{\theta}\hat{r}\hat{\theta}}\right)_W=\frac{1}{144}\left(a^{-4}\hat{p}_a^2+4ia^{-5}\hat{p}_a-5a^{-6}\right).
\label{curvature weyl ordering}\end{aligned}
\end{equation}

Now the local expectation value of the components of the Riemann tensor are obtained by the relation
\begin{equation}
\left<\hat{R}_{\hat{a}\hat{b}\hat{c}\hat{d}}\right>_L=\text{Re}\left[\frac{\Psi^{\ast}\left(\hat{R}_{\hat{a}\hat{b}\hat{c}\hat{d}}\right)_W\Psi}{\Psi^{\ast}\Psi}\right].
\label{lev components}
\end{equation}

We have
\begin{equation}
\Psi^{\ast}\left(\frac{\hat{p}_a^2}{a^4}\right)_W\Psi=\Theta e^{-iS}\left[a^{-4}\hat{p}_a^2+4ia^{-5}\hat{p}_a-5a^{-6}\right]\Theta e^{iS}.
\label{pa a weyl}
\end{equation}

But
\begin{equation}\begin{aligned}
\left(\Theta e^{-iS}\right)a^{-4}\hat{p}_a^2\left(\Theta e^{iS}\right)&=-\left(\Theta e^{-iS}\right)a^{-4}\frac{\partial^2}{\partial a^2}\left(\Theta e^{iS}\right)\\
&=-\Theta a^{-4}\left[\frac{\partial^2\Theta}{\partial a^2}+2i\frac{\partial\Theta}{\partial a}\frac{\partial S}{\partial a}+i\Theta\frac{\partial^2 S}{\partial a^2}-\Theta \left(\frac{\partial S}{\partial a}\right)^2\right]
\label{relations1}\end{aligned}
\end{equation}
and
\begin{equation}
\left(\Theta e^{-iS}\right)4ia^{-5}\hat{p}_a\left(\Theta e^{iS}\right)=\left(\Theta e^{-iS}\right)4a^{-5}\frac{\partial}{\partial a}\left(\Theta e^{iS}\right)=4\Theta a^{-5}\left(\frac{\partial \Theta}{\partial a}+i\Theta\frac{\partial S}{\partial a}\right).
\label{relations2}
\end{equation}
Therefore
\begin{equation}
\left<\frac{\hat{p}_a^2}{a^{4}}\right>_L=a^{-4}\left(\frac{\partial S}{\partial a}\right)^2-\frac{a^{-4}}{\Theta}\frac{\partial^2\Theta}{\partial a ^2}+\frac{4a^{-5}}{\Theta}\frac{\partial \Theta}{\partial a}-5a^{-6}.
\label{pa^2/a^{-4} weyl}
\end{equation}

The quantum mechanical potential \cite{bohm} in the de Broglie-Bohm interpretation is obtained through the equation
\begin{equation}
Q=-\frac{1}{24 \Theta}\frac{\partial^2 \Theta}{\partial a^2}=\frac{3\sigma}{\sigma^2t^2+36}-\frac{216\sigma^2a^2}{(\sigma^2t^2+36)^2}.
\label{quantum potential}
\end{equation}

The time rate of the momentum $p_a$ has the following value \cite{holland}
\begin{equation}
\dot{p}_a=-\frac{\partial}{\partial a}(V+Q),
\label{time rate momentum}
\end{equation}
where $V$ is the classical potential. In our case $V=0$ so that
\begin{equation}
\dot{p}_a=\frac{432\sigma^2a}{(\sigma^2t^2+36)^2}.
\label{time rate momentum 2}
\end{equation}

From Eqs. (\ref{theta and S}) and (\ref{pa^2/a^{-4} weyl}) we have
\begin{equation}\begin{aligned}
\left<\frac{\hat{p}_a^2}{a^{4}}\right>_L&=\frac{1}{a(t)^4}\left(\frac{12\sigma^2ta(t)}{36+\sigma^2t^2}\right)^2-\frac{1}{a(t)^4}\left[\frac{-72\sigma}{36+\sigma^2t^2}+\left(\frac{72\sigma a(t)}{36+\sigma^2t^2}\right)^2\right]+\frac{4}{a(t)^5}\left(\frac{-72\sigma a(t)}{36+\sigma^2t^2}\right)-\frac{5}{a(t)^6}\\
&=\frac{-5+72\sigma a_0^2[-3+2\sigma(-36+t^2\sigma^2)a_0^2]}{(36+t^2\sigma^2)^3a_0^6}.
\label{P2}\end{aligned}
\end{equation}

Substituting Eqs. (\ref{time rate momentum 2}) and (\ref{P2}) into Eq. (\ref{components momentum}), we have
\begin{equation}\begin{aligned}
&\left<\hat{R}_{\hat{t}\hat{r}\hat{t}\hat{r}}\right>_L=\frac{-5+72\sigma a_0^2 [-3+2\sigma(-72+\sigma^2t^2)a_0^2]}{144(36+\sigma^2t^2)^3a_0^6}\\
&\left<\hat{R}_{\hat{r}\hat{\theta}\hat{r}\hat{\theta}}\right>_L=\frac{-5+72\sigma a_0^2[-3+2\sigma(-36+\sigma^2t^2)a_0^2]}{144(36+\sigma^2t^2)^3a_0^6}.
\label{components final}\end{aligned}
\end{equation}

\begin{figure}[!htb]
\centering
\includegraphics[scale=1]{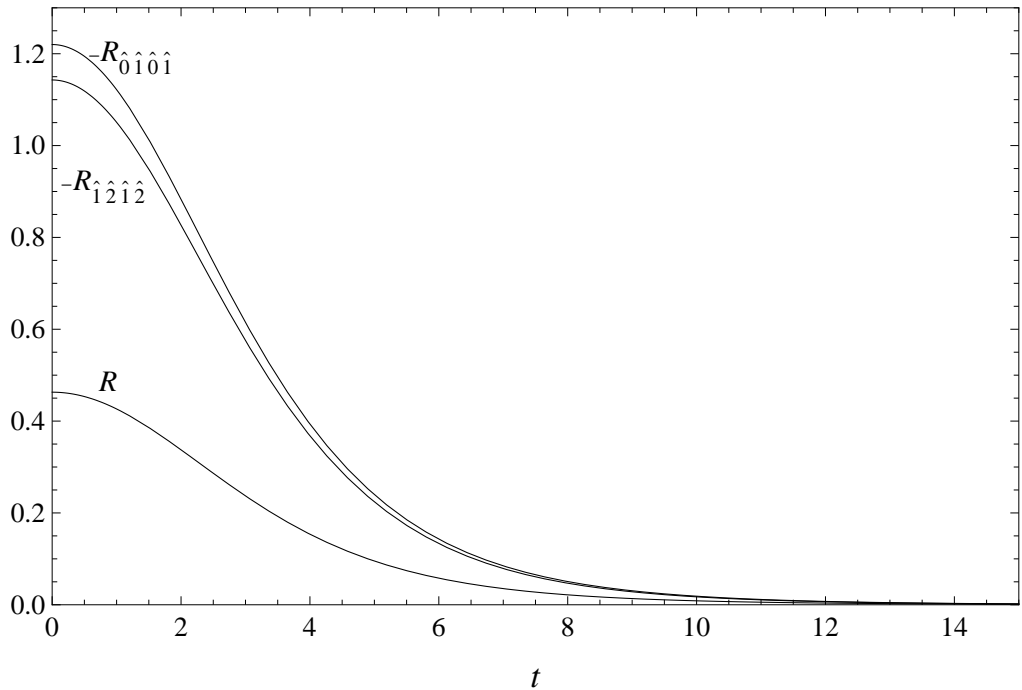}
\caption{The local expectation value of the non-null curvature components in the tetrad basis and of the Ricci scalar. They are perfectly for all values of $t$. Here we have chosen $\sigma=1$ and $a_0=0.1$.}
\label{graphics components}
\end{figure}

The graphics of $\left<\hat{R}_{\hat{t}\hat{r}\hat{t}\hat{r}}\right>_L$ and $\left<\hat{R}_{\hat{r}\hat{\theta}\hat{r}\hat{\theta}}\right>_L$ are shown in Fig. \ref{graphics components}. Note that they are perfect regular for all time $t$, indicating the absence of a big-bang singularity in the quantum cosmological model. If they were classical quantities, every curvature scalar could be constructed from them. Of course this is not true in our case (for example, if we would like to quantize the Kretschmann scalar, we would face a factor ordering problem involving a term like $p^{4}_{a}/a^{8}$, see Ref. \cite{monerat}), but it indicates that all the curvature scalars are perfectly regular.

The Ricci scalar is related to the independent components of the curvature tensor by the equation
\begin{equation}
R=-6R_{\hat{t}\hat{r}\hat{t}\hat{r}}+6R_{\hat{r}\hat{\theta}\hat{r}\hat{\theta}}=\frac{\dot{p}_a}{2a^3}=\frac{216\sigma^2}{a^2(\sigma^2t^2+36)^2}.
\label{Ricci scalar}
\end{equation}
Its graphic, Fig. \ref{graphics components}, is perfectly regular like in Ref. \cite{monerat}, giving further evidence of the nonsingular character of the quantum cosmological model.

In the tetrad basis, the tensor $T^{\mu}_{\phantom{\mu}\nu}$ is in the diagonal form
\begin{equation}
T^{\hat{\mu}}_{\phantom{\hat{\mu}}\hat{\nu}}=\text{diag}(-\rho,p,p,p).
\label{energy-momentum}
\end{equation}
Using the Einstein equation
\begin{equation}
T^{\mu}_{\phantom{\mu}\nu}=\frac{1}{8\pi}\left(R^{\mu}_{\phantom{\mu}\nu}-\frac{1}{2}\delta^{\mu}_{\nu}R\right)
\label{einstein equation}
\end{equation}
we have
\begin{equation}\begin{aligned}
&T^{\hat{0}}_{\phantom{\hat{0}}\hat{0}}=-\rho=-\frac{3}{8\pi}R_{\hat{r}\hat{\theta}\hat{r}\hat{\theta}},\\
&T^{\hat{i}}_{\phantom{\hat{i}}\hat{i}}=p=\frac{1}{8\pi}\left(2R_{\hat{t}\hat{r}\hat{t}\hat{r}}-R_{\hat{r}\hat{\theta}\hat{r}\hat{\theta}}\right).
\label{rho and p}\end{aligned}
\end{equation}

\begin{figure}[!htb]
\centering
\includegraphics[scale=1]{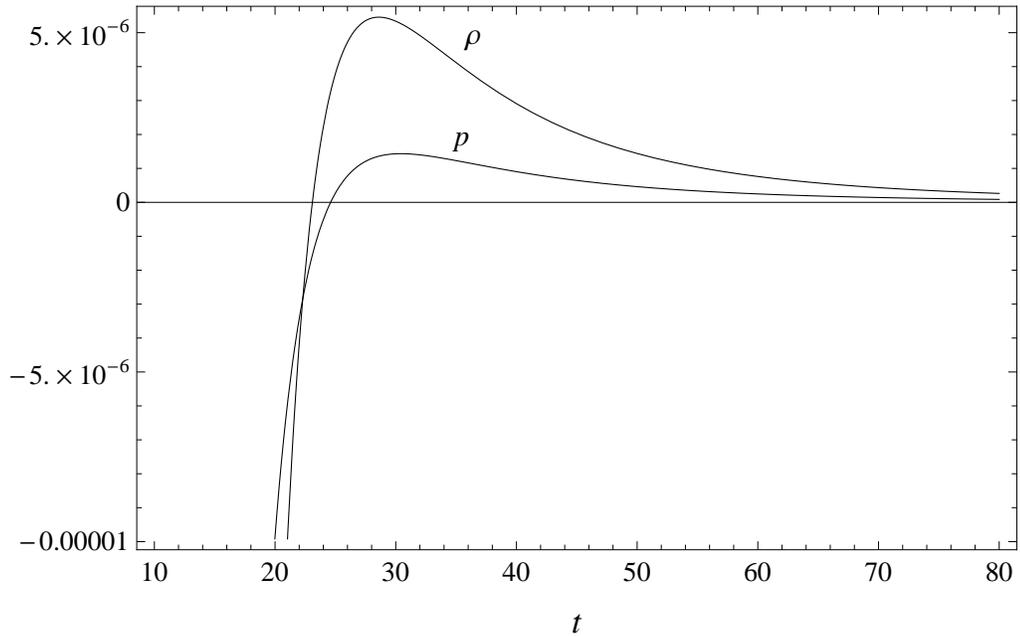}
\caption{The local expectation value of the energy density and pressure for $\sigma=1$ and $a_0=0.1$.}
\label{graphics rho and p}
\end{figure}

Now we can plot the graphics of $\left<\hat{\rho}\right>_L$ and $\left<\hat{p}\right>_L$ (Fig. \ref{graphics rho and p}). At the quantum level both are negative and turn positive positive as quantum effects become negligible. They are also regular for all times $t$, including the beginning of times $t=0$.

\begin{figure}[!htb]
\centering
\includegraphics[scale=1]{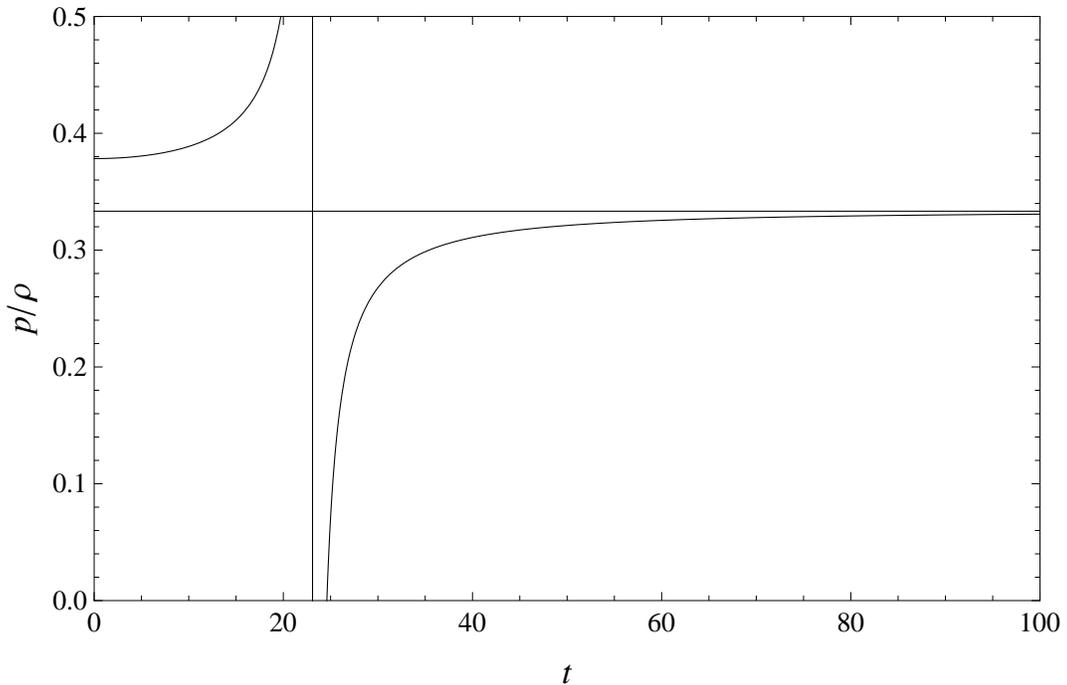}
\caption{The local expectation value of $p/\rho$ for $\sigma=1$ and $a_0=0.1$. The function explodes in $t$ between $20$ and $25$ because $\rho$ becomes zero at this point.}
\label{equation of state}
\end{figure}

We can also study $\left<\hat{p}\right>_L/\left<\hat{\rho}\right>_L$ in order to see if the equation of state is respected during the evolution of the Universe. In Fig. \ref{equation of state} we note that in the classical limit $\left<\hat{p}\right>_L/\left<\hat{\rho}\right>_L\to 1/3$ as expected. This shows the consistency of the model, since we expect that as $t$ becomes large, classical aspects begin to appear.

We now turn to analyze the breakdown in the energy conditions. There are three types of physical reasonable energy conditions to be considered \cite{hawking}. The {\it weak energy  condition} states that $T_{\mu\nu}W^{\mu}W^{\nu}\geq 0$ for any timelike vector $W^{\mu}$. For an energy-momentum tensor expressed in the form (\ref{energy-momentum}), this will be true if and only if $\rho\geq0$ and $\rho+p\geq0$. The {\it dominant energy condition} says that for every timelike vector $W^{\mu}$, $T_{\mu\nu}W^{\mu}W^{\nu}\geq0$ and $T^{\mu\nu}W_{\nu}$ is a nonspacelike vector. This holds if $\rho\geq0$ and $\rho\geq \left|p\right|$. Finally the most common energy condition, called {\it strong energy condition}, states that $R_{\mu\nu}W^{\mu}W^{\nu}\geq0$ for every timelike vector $W^{\mu}$. According to the Einstein equation this is equivalent to saying that the energy-momentum tensor satisfies $T_{\mu\nu}W^{\mu}W^{\nu}\geq \frac{1}{2}W^{\mu}W^{\mu}T$ and this will be true for the energy-momentum tensor (\ref{energy-momentum}) if $\rho+p\geq0$ and $\rho+3p\geq0$. These conditions, along with some simple restrictions on the spacetime manifold \cite{tipler 2}, such as certain reasonable initial conditions (the existence of trapped surfaces or the existence of a spacelike hypersurface) and restrictions on the causal structure (the existence of a Cauchy surface or the absence of closed timelike curves), give rise to the Hawking-Penrose singularity theorems. They state that under these conditions singularities must occur in the spacetime. 

\begin{figure}[!htb]
\centering
\includegraphics[scale=1]{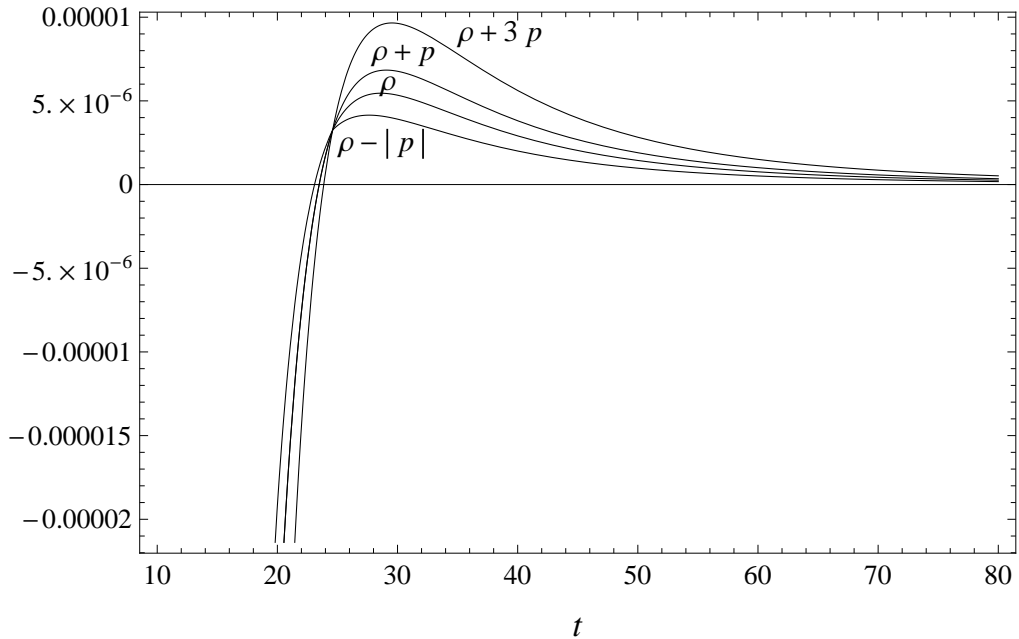}
\caption{The energy conditions for $\sigma=1$ and $a_0=0.1$. We see that every one of them are violated at the quantum level.}
\label{graphics energy conditions}
\end{figure}

In Fig. \ref{graphics energy conditions} we can see the local expectation value of the various relations between $\rho$ and $p$ implied by the energy conditions. We note that they are all violated at the quantum level as we would expect, since the singularity has been removed, and becomes valid in the classical limit, showing the consistency of the model. 

Now we will show that these results are inherent of the quantum model, not of a particular choice of the wave packet. For this we will do the same analysis using a different wave packet satisfying the Dirichlet boundary condition, i.e., $\Psi(0,t)=0$ [$\beta=\infty$ in Eq. (\ref{boundary conditions})]. 

The Green function for the Dirichlet boundary condition is given by
\begin{equation}
G^{\beta=\infty}(a,a',t)=G(a,a',t)-G(a,a',t).
\end{equation}

If we start with a normalized wave packet
\begin{equation}
\Psi(a,0)=\left(\frac{128\sigma^3}{\pi}\right)^{1/4}ae^{-\sigma a^2}
\end{equation}
we have
\begin{equation}\begin{aligned}
\Psi(a,t)&=\int_{0}^{\infty}{G^{\beta=\infty}(a,a't)\Psi(a',0)da'}=\int_{-\infty}^{\infty}{G(a,a't)\Psi(a',0)da'}\\&=\left(\frac{128\sigma^3}{\pi}\right)\frac{(216i)^{1/2}}{(\sigma t-6i)^{3/2}}a\exp{\left(\frac{6i\sigma a^2}{\sigma t-6i}\right)}.
\end{aligned}\end{equation}
The polar decomposition of $\Psi$ has the form
\begin{displaymath}
\Psi=\Theta e^{iS},
\end{displaymath}
where
\begin{equation}\begin{aligned}
&S=\frac{6\sigma^2ta^2}{\sigma^2t^2+36}+f_0(t),\\
&\Theta=g_0(t)a\exp{\left(\frac{-36\sigma a^2}{\sigma^2t^2+36}\right)}.
\end{aligned}\end{equation}

The Bohmian trajectory for the scale factor is the same as the previous case, i.e., 
\begin{displaymath}
a(t)=a_0\sqrt{36+\sigma^2t^2}.
\end{displaymath}

The local expectation value for the quantity $\hat{p}_a^2/a^4$ is now given by
\begin{equation}\begin{aligned}
\left<\frac{\hat{p}_a^2}{a^{4}}\right>_L=&\frac{1}{a(t)^4}\left(\frac{12\sigma^2ta(t)}{36+\sigma^2t^2}\right)^2-\frac{1}{a(t)^4}\frac{1}{a(t)}\left[-\frac{144\sigma a(t)}{26+\sigma^2t^2}+\left(1-\frac{72\sigma a(t)^2}{\sigma^2t^2+36}\right)\left(-\frac{72\sigma a(t)}{\sigma^2t^2+36}\right)\right]\\&+\frac{4}{a(t)^5}\frac{1}{a(t)}\left(1-\frac{72\sigma a(t)^2}{36+\sigma^2t^2}\right)-\frac{5}{a(t)^6}=\frac{-1+72\sigma a_0^2[-1+2\sigma(-36+t^2\sigma^2)a_0^2]}{(36+t^2\sigma^2)^3a_0^6}
\end{aligned}
\end{equation}

The local expectation values of the components of the curvature tensor in the tetrad basis are given by
\begin{equation}\begin{aligned}
&\left<\hat{R}_{\hat{t}\hat{r}\hat{t}\hat{r}}\right>_L=\frac{-1+72\sigma a_0^2 [-1+2\sigma(-72+\sigma^2t^2)a_0^2]}{144(36+\sigma^2t^2)^3a_0^6}\\
&\left<\hat{R}_{\hat{r}\hat{\theta}\hat{r}\hat{\theta}}\right>_L=\frac{-1+72\sigma a_0^2[-1+2\sigma(-36+\sigma^2t^2)a_0^2]}{144(36+\sigma^2t^2)^3a_0^6}.
\end{aligned}
\end{equation}

\begin{figure}[!htb]
\centering
\includegraphics[scale=1]{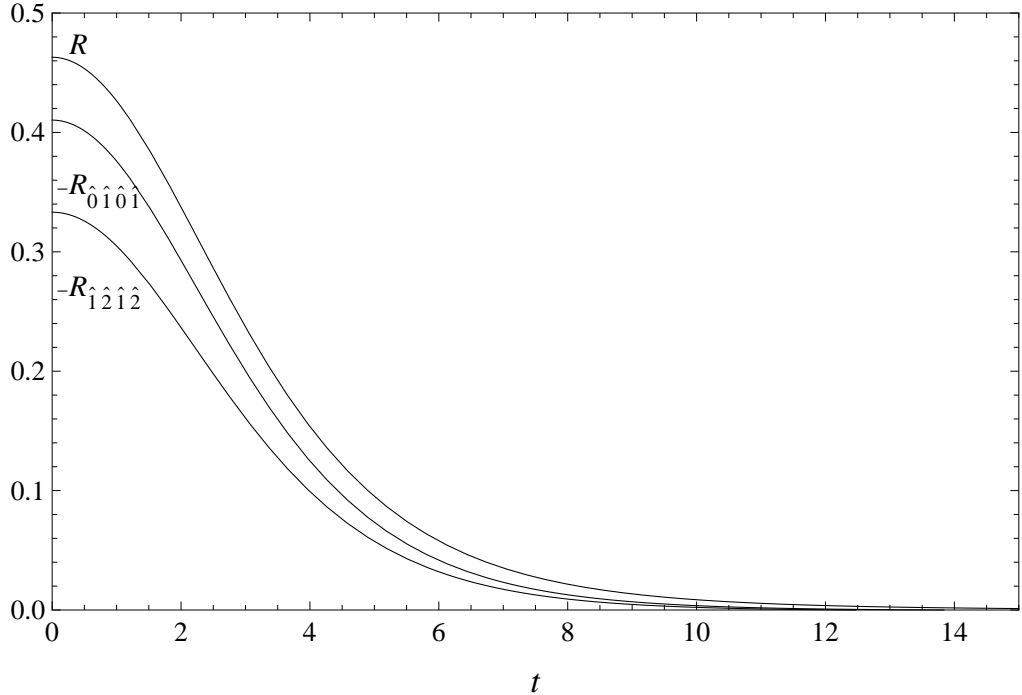}
\caption{The local expectation value of the non-null curvature components in the tetrad basis and of the Ricci scalar for a wave packet satisfying the Dirichlet boundary condition for $\sigma=1$ and $a_0=0.1$.}
\label{figura5}
\end{figure}

\begin{figure}[!htb]
\centering
\includegraphics[scale=1]{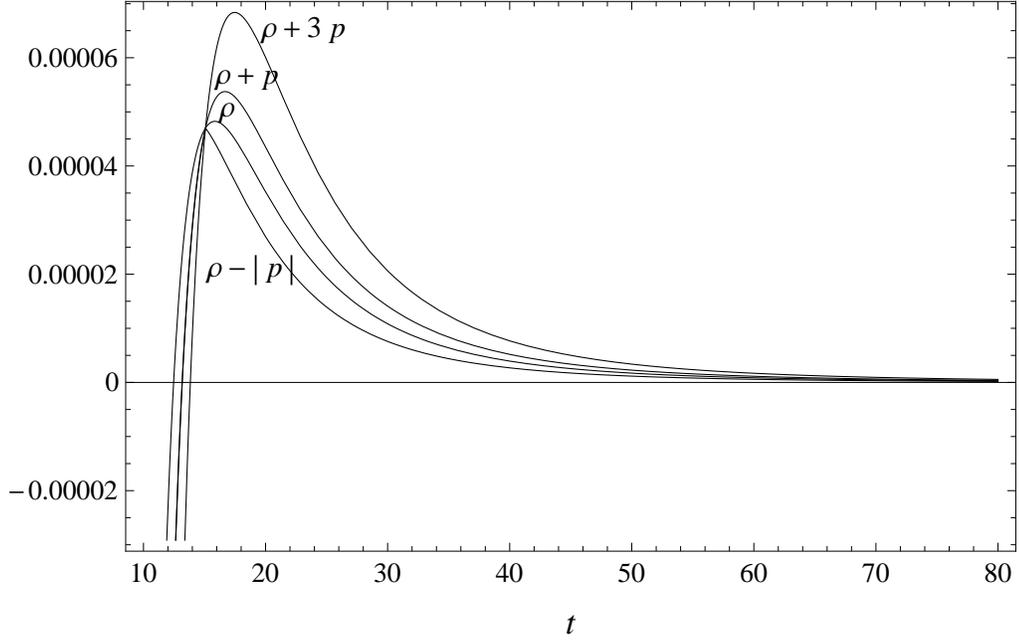}
\caption{The energy conditions for a wave packet satisfying $\Psi(a,0)=0$ for $\sigma=1$ and $a_0=0.1$.}
\label{figura6}
\end{figure}

In what follows we will show the graphics of $\left<\hat{R}_{\hat{t}\hat{r}\hat{t}\hat{r}}\right>_L$, $\left<\hat{R}_{\hat{r}\hat{\theta}\hat{r}\hat{\theta}}\right>_L$ and $\left<R\right>_L$ in Fig. \ref{figura5} and the relations between $\rho$ and $p$  implied by the energy conditions in Fig. \ref{figura6}. As we can see, the graphics in this case are very similar to those showed previously.

The other graphics representing $\rho$, $p$ and $p/\rho$ and  we did not show here are very similar to those showed previously too.

\section{Independence of the nonsingular character of the quantum cosmological model with respect to the wave packet}
\label{section 4}
In this section we calculate the expectation value of the scale factor in the many-worlds interpretation of quantum mechanics for several wave packets representing the wave function of the Universe. We want to speculate as to whether the nonsingular character of the quantum Universe is associated with the simple wave packets found in the literature so far. 

The wave packets we will use in this section will be obtained in two different ways. The first by using the Green function [Eq. (\ref{green function})] with the boundary condition $\Psi'(0,t)=0$. The second is through the eingenfunctions \cite{alvarenga}
\begin{equation}
\Psi_E(a,t)=e^{-iEt}\sqrt{a}J_{1/2}\left(\frac{\sqrt{96E}}{2}a\right)
\label{eingeinfunctions}
\end{equation}
of Eq. (\ref{wheeler-dewitt equation}). The superposition
\begin{equation}
\Psi(a,t)=\int_{0}^{\infty}{A(E)\Psi_E(a,t)dE}
\label{superposition}
\end{equation}
implies the boundary condition $\Psi(0,t)=0$.

The first wave packet we use is given in Eq. (\ref{wave packet with green function}). The expectation value of the scale factor for such wave function was calculated in Ref. \cite{lemos} and is given by
\begin{equation}
<a(t)>=\frac{1}{12}\sqrt{\frac{2}{\pi\sigma}}\sqrt{\sigma^2t^2+36}.
\label{wave 1}
\end{equation}

Now, we will take the initial wave packets of the form
\begin{equation}
\Psi(a,0)=a^{n}e^{-\gamma a^2}\;\;\;\;\; \gamma>0,
\label{initial waves}
\end{equation}
n even. Then we have
\begin{equation}
\Psi(a,t)=\left(\frac{6}{\pi i t}\right)^{1/2}\int_{-\infty}^{\infty}{e^{\frac{6i}{t}(a-a')^2}a'^n e^{-\gamma a'^2}da'}.
\label{psi t}
\end{equation}

This integral can be performed with the help of Ref. \cite{gradshteyn} and gives
\begin{equation}
\Psi(a,t)=\alpha_n(\gamma,t)\exp{\left(-\frac{36a^2\gamma t^2}{\gamma^2t^4+36t^2}\right)}\exp{\left[i\left(-\frac{216a^2t}{\gamma^2t^4+36t^2}+\frac{6a^2}{t}\right)\right]}H_n\left(\frac{6a}{(\gamma t^2-6i)^{1/2}}\right),
\label{psi after integral}
\end{equation}
where
\begin{equation}
\alpha_n(\gamma,t)=6^{1/2}t^{n/2}(i)^{-(n+1)/2}2^{-n}(\gamma t-6i)^{-(n+1)/2}.
\label{alpha}
\end{equation}
and $H_n(x)$ is the Hermite polynomial of order $n$. Let us consider the case $n=2$ and $n=4$. First let us take $\Psi(a,0)=a^2e^{-\gamma a^2}$. In this case
\begin{equation}
\Psi(a,t)=\alpha_2(\gamma,t)e^{-\frac{\xi}{2} a^2}e^{i\nu}(\eta a^2-2),
\label{a^2}
\end{equation}
\begin{equation}\begin{aligned}
&\xi=-\frac{72\gamma t^2}{\gamma^2t^4+36t^2},\\
&\nu=-\frac{216a^2t}{\gamma^2t^4+36t^2}+\frac{6a^2}{t},\\
&\eta=\frac{144}{\gamma t^2-6it}.
\end{aligned}
\label{parameters values}
\end{equation}

The expectation value of the scale factor $a$ for this wave packet is given by
\begin{equation}\begin{aligned}
\left<a\right>(t)&=\frac{\int_{0}^{\infty}{\Psi(a,t)^{\ast}a\Psi(a,t)}da}{\int_{0}^{\infty}{\Psi(a,t)^{\ast}\Psi(a,t)}da}=\frac{\int_{0}^{\infty}{\left|\alpha(\gamma,t)\right|e^{-\xi a^2}\left[\left|\eta\right|^2a^5-2\text{Re}(\eta)a^3+4a\right]}da}{\int_{0}^{\infty}{\left|\alpha(\gamma,t)\right|e^{-\xi a^2}\left[\left|\eta\right|^2a^4-2\text{Re}(\eta)a^2+4\right]}da}\\
&=\frac{\sqrt{\frac{\pi}{2}}(72+t^2\gamma^2)}{9\sqrt{\gamma(36+t^2\gamma^2)}}.
\label{expectation a^2}\end{aligned}
\end{equation}

We note that the expectation value of the scale factor is nonzero for all times, showing that the universe is nonsingular. Now let us take $\Psi(a,0)=a^4e^{-\gamma a^2}$. In this case
\begin{equation}
\Psi(a,t)=\alpha_4(\gamma,t)e^{-\frac{\xi}{2} a^2}e^{i\nu}(\eta a^4-\varsigma a^2+12),
\label{a^4}
\end{equation}
where
\begin{equation}\begin{aligned}
&\eta=\frac{20736}{(\gamma t^2-6it)^2},\\
&\varsigma=\frac{1728}{\gamma t^2-6it}.
\label{parameters value 2}\end{aligned}
\end{equation}
Then we have
\begin{equation}
\left<a\right>(t)=\frac{4\sqrt{\frac{2}{\pi}}(3456+144t^2\gamma^2+t^4\gamma^4)}{35\sqrt{\gamma}(36+t^2\gamma^2)^{3/2}}.
\label{expectation a^4}
\end{equation}
Again the expectation value of the scale factor is nonzero for all values of $t$. Note that on both cases the asymptotic behavior of the scale factor is $\left<a\right>(t)\propto t$, like its classical analogue, since we are working with the gauge choice $N=a$ so that the cosmological time $\tau$ (in the asymptotic behavior) is given by
\begin{equation}
d\tau=tdt\Rightarrow\tau\propto t^2\Rightarrow \left<a\right>(\tau)\propto \tau^{1/2}.
\label{cosmological time}
\end{equation}

Let us now take more elaborate wave packets. Consider $A(E)=e^{-E}$ in Eq. (\ref{superposition}). The integral
\begin{equation}
\Psi(a,t)=\sqrt{a}\int_{0}^{\infty}{e^{-(1+it)E}J_{1/2}(\frac{\sqrt{96E}}{2}a)dE}
\label{integral a^{-gamma E}}
\end{equation}
can be solved analytically \cite{gradshteyn}, and we get
\begin{equation}
\Psi(a,t)=\frac{\sqrt{96}}{8}\sqrt{\frac{\pi}{(\gamma+it)^3}}a^{3/2}e^{\frac{3a^2it}{\gamma^2+t^2}}e^{-\frac{3\gamma a^2}{\gamma^2+t^2}}\left[I_{-1/4}\left(\frac{3a^2}{\gamma+it}\right)-I_{3/4}\left(\frac{3a^2}{\gamma+it}\right)\right].
\label{psi e^{-E}}
\end{equation} 

\begin{figure}[!htb]
\centering
\includegraphics[scale=1]{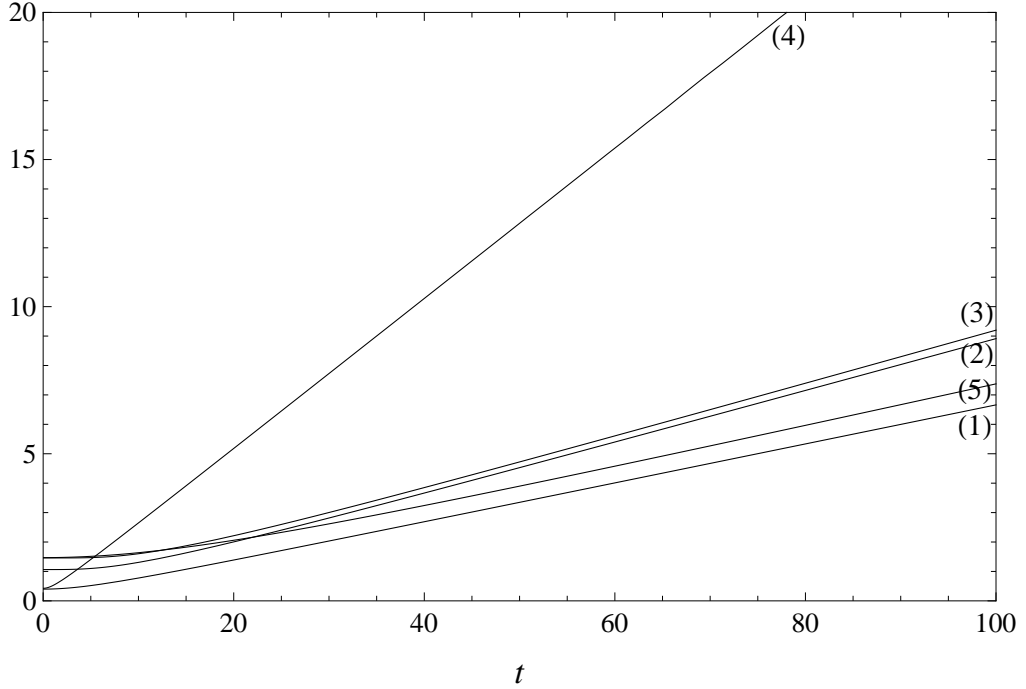}
\caption{ The expectation value of the scale factor for $(1)$ $\Psi(a,0)=e^{-a^2}$, $(2)$ $\Psi(a,0)=a^2e^{-a^2}$, $(3)$ $\Psi(a,0)=a^4e^{-a^2}$, $(4)$ $A(E)=e^{-E}$ and $(5)$ $A(r)=e^{-r^2}I_{1/2}(r)$. We see that they have qualitatively the same behavior, i.e., asymptotically they behave as straight lines which pass trough the origin}. 
\label{FRW6}
\end{figure}

We were unable to   solve analytically the expectation value of the scale factor for this case, but we solved numerically. The result is shown in Fig. \ref{FRW6}.

If we now define $r=\frac{\sqrt{96}}{2}$ in Eq. (\ref{eingeinfunctions}) we will have, instead of Eq. (\ref{superposition}),
\begin{equation}
\Psi(a,t)=\frac{\sqrt{a}}{12}\int_{0}^{\infty}{r e^{\frac{-ir^2 t}{24}}J_{1/2}(ra)A(r)dr}.
\label{superposition 2}
\end{equation}

Consider $A(r)=e^{-\gamma r^2}I_{1/2}(\beta r)$. Then, the integral (\ref{superposition 2}) can be solved \cite{gradshteyn} and results in
\begin{equation}\begin{aligned}
\Psi(a,t)=&\frac{\sqrt{a}}{24\left(\gamma+\frac{it}{24}\right)}\exp{\left(-i\frac{\beta^2t}{96\left(\gamma^2+\frac{t^2}{24^2}\right)}\right)}\exp{\left(i\frac{a^2t}{96\left(\gamma^2+\frac{t^2}{24^2}\right)}\right)}\times\\
&\times\exp{\left(\frac{\beta^2\gamma}{4\left(\gamma^2+\frac{t^2}{24^2}\right)}\right)}\exp{\left(-\frac{\gamma a^2}{4\left(\gamma^2+\frac{t^2}{24^2}\right)}\right)}J_{1/2}\left(\frac{\beta a}{2(\gamma+\frac{it}{24})}\right).
\label{wave packet with I}\end{aligned}
\end{equation}
This wave packet respects the boundary condition $\Psi(0,t)=0$, and the initial wave packet is given by
\begin{equation}
\Psi(a,0)=\frac{\sqrt{a}}{24\gamma}e^{\frac{\beta^2\gamma}{4\gamma^2}}e^{-\frac{\gamma a^2}{4\gamma^2}}J_{1/2}\left(\frac{\beta a}{2\gamma}\right).
\label{initial wave I}
\end{equation}

We could not find either an analytic expression for the expectation value of the scale factor, but we have shown the numerical result in Fig. \ref{FRW6}.

Fig. \ref{FRW6} shows the expectation value of the scale factor for $(1)$ $\Psi(a,0)=e^{-a^2}$, $(2)$ $\Psi(a,0)=a^2e^{-a^2}$, $(3)$ $\Psi(a,0)=a^4e^{-a^2}$, $(4)$ $A(E)=e^{-E}$ and $(5)$ $A(r)=e^{-r^2}I_{1/2}(r)$. We see that they have qualitatively the same behavior, every one of them growing linearly proportional to $t$ as $t\to\infty$. The main point here is that asymptotically the graphics behave as straight lines which pass trough the origin. In principle the proportionality constants  need not be the same. In every case the singularity has been excluded, indicating that this is a property of the introduction quantum cosmology, not of a particular wave packet.

\section{Recovering the Big Bang by the introduction of a nonnormalizable wave packet}
\label{section 5}

It has been argued that a nonnormalizable wave function is unavoidable in quantum cosmology \cite{tipler}. With such wave packet we cannot find the expectation value of the scale factor, but we can still find the trajectory in the de Broglie-Bohm interpretation. The role of the wave function in the ontological interpretation of quantum mechanics is to provide a wave guidance [Eq. (\ref{momentum bohmian})] for the dynamical variables, and this clearly does not require normalization \cite{shtanov}. Square integrability of the wave function is necessary only when probability takes place. So it is valid to work with nonnormalizable wave functions if we are looking for the trajectory of a dynamical variable. Now we will show that the singular character of the scale factor can be recovered with the introduction of a nonnormalizable solution of Eq. (\ref{wheeler-dewitt equation}) in the de Broglie-Bohm interpretation of quantum mechanics. 

First, let us assume that the wave function satisfies the boundary condition $\Psi(0,t)=0$. This boundary condition was defended by  DeWitt \cite{dewitt} because it keeps wave packets away from the singularity. Now we take $A(E)=J_{1/2}(\beta\sqrt{E})$ in Eq. (\ref{superposition}) and we have
\begin{equation}
\Psi(a,t)=\sqrt{a}\int_{0}^{\infty}{e^{-iEt}J_{1/2}\left(\frac{\sqrt{96}a}{2}\sqrt{E}\right)J_{1/2}(\beta\sqrt{E})dE}.
\label{bessel b root E}
\end{equation}

The above integral can be analytically solved with the help of Ref. \cite {gradshteyn}. The result is
\begin{equation}
\Psi(a,t)=\sqrt{a}\frac{1}{t}J_{1/2}\left(\frac{\sqrt{96}a\beta}{4t}\right)\exp{\left[i\left(\frac{24a^2+\beta^2}{4t}-\frac{3\pi}{4}\right)\right]}.
\label{solution bessel b root E}
\end{equation}

It is already separated in the polar form $\Theta e^{iS}$, so we have
\begin{equation}
S=\frac{24a^2+\beta^2}{4t}-\frac{3\pi}{4}.
\label{S}
\end{equation} 

By the Eq. (\ref{momentum bohmian}) we have
\begin{equation}
12\dot{a}=\frac{12a}{t}\Rightarrow a(t)=a_0 t.
\label{momentum bessel}
\end{equation}

Note that the above Bohmian trajectory of the scale factor recovers the classical behavior $a(\tau)\propto\tau^{1/2}$ for the radiation dominated era, where $\tau$ is the cosmological time.

Now, following the quantization procedure discussed by Tipler in Ref. \cite{tipler}, we require that, like the classical case in which all solutions pass through the singularity $a=0$ in $t=0$, all quantum universes do the same. Then we pick $\Psi(a,0)=\delta(a)$ as the initial wave packet. It is easy to show that for such an initial wave packet, the DeWitt boundary condition gives rise to a null wave function for the Universe. So we take the boundary condition used by Tipler in \cite{tipler} $\Psi'(0,t)=0$. Then we have
\begin{equation}
\Psi(a,t)=\int_{-\infty}^{\infty}{G(a,a',t)\delta(a')da'}=G(a,0,t)=\left(\frac{6}{\pi i t}\right)^{1/2}\exp{\left(\frac{6i}{t}a^2\right)}.
\label{solution tipler}
\end{equation}

This is what Tipler called the Green function of the universe. Note that in this case we have
\begin{equation}
S=\frac{6a^2}{t},
\label{S in tipler's case}
\end{equation}
so that
\begin{equation}
12\dot{a}=\frac{12a}{t}\Rightarrow a(t)=a_0t
\label{partial S}
\end{equation}
like in the previous case.

It is worth noting that, as observed by Lemos in \cite{lemos}, these boundary conditions are somewhat arbitrary for the wave packets studied in this section. This is because, since these functions are not square-integrable, they do not satisfy the condition for the Hermiticity of the operator $\hat{H}=-\frac{\partial^2}{\partial a^2}$
\begin{equation}
\left(\Psi^{\ast}\frac{d\Psi}{da}-\frac{d\Psi^{\ast}}{da}\Psi\right)(\infty)=\left(\Psi^{\ast}\frac{d\Psi}{da}-\frac{d\Psi^{\ast}}{da}\Psi\right)(0),
\label{condition hermiticity}
\end{equation}
so they are not in the domain of functions where $\hat{H}$ is Hermitian. So equation (\ref{boundary conditions}) is not a condition for self-adjointness of  the operator $\hat{H}$ in this case.

\section{conclusions}
\label{section 6}
By the quantization of the components of the curvature tensor in the tetrad basis and with the use of a simple wave packet for the Universe we were able to take the local expectation value of these components in the de Broglie-Bohm interpretation of quantum mechanics. We then showed graphically that these quantities are perfectly regular for all times, giving one further evidence that the universe is nonsingular with the introduction of quantum cosmology, like in Ref. \cite{monerat}. We also took the local expectation value of the energy density $\rho$ and the pressure $p$ of the fluid and showed that the energy conditions are violated in the quantum era as  expected, since under the validity of these conditions the Universe is undoubtedly singular. We also found the consistency of the theory by showing that the classical equation of state relating $\rho$ and $p$ is recovered in the classical limit and that the classical energy conditions are again observed when quantum mechanics becomes unimportant.

We also find the expectation  value of the scale factor in the many-worlds interpretation for several wave packets representing the wave function of the Universe. We showed graphically that for every wave packet, the qualitative behavior of the scale factor is the same, every one of them tending to the classical expression $a(t)\propto t$ ($a(\tau)\propto \tau^{1/2}$). As we said before, it has been argued that nonnormalizable wave packets are unavoidable to quantum cosmology, so we took two nonnormalizable wave packets representing wave functions for the Universe and showed that for these wave packets, the Bohmian trajectory of the scale factor is singular like its classical analogue, so we could speculate that if the big-bang singularity really existed, the wave function of the universe is in fact nonnormalizable.

\acknowledgements 
J.P.M. Pitelli thanks FAPESP for financial support. P.S. Letelier aknowledges FAPESP and CNPq for partial financial support. Discussions with N. Lemos and J. Fabris are also acknowledged.


\end{document}